\documentclass[epj]{webofc}
\usepackage[utf8]{inputenc}
\usepackage[varg]{txfonts}   % Web of Conferences font
\usepackage{booktabs}
\usepackage{xcolor}
\definecolor{darkred}{rgb}{0.4,0.0,0.0}
\definecolor{darkgreen}{rgb}{0.0,0.4,0.0}
\definecolor{darkblue}{rgb}{0.0,0.0,0.4}
\usepackage[bookmarks,linktocpage,colorlinks,
    linkcolor = darkred,
    urlcolor  = darkblue,
    citecolor = darkgreen]{hyperref}
\usepackage{subfigure}
\wocname{EPJ Web of Conferences}
\woctitle{Lattice2017}
\newcommand{\be}{\begin{equation}}
\newcommand{\ee}{\end{equation}}
\newcommand{\bea}{\begin{eqnarray}}
\newcommand{\eea}{\end{eqnarray}}
\begin{document}
%%%%%%%%%%%%%%%%%%%%%%%%%%%%%%%%%%%%%%%%%%%%%%%%%%%%%%%%%%%%%%%%%%%%%%%%%%%%%
%
\selectlanguage{english}
%----------------------------------------------------------------------------
\title{%
HQE parameters from unquenched lattice data on pseudoscalar and vector heavy-light meson masses
}
%----------------------------------------------------------------------------
\author{%
\firstname{Paolo} \lastname{Gambino}\inst{1}\and
\firstname{Aurora} \lastname{Melis}\inst{2} \fnsep\thanks{Speaker, \email{aurora.melis@ific.uv.es}. Acknowledges financial support from ``La Caixa-Severo Ochoa'' scholarship.}\and
\firstname{Silvano}  \lastname{Simula}\inst{3}
% etc.
}
%----------------------------------------------------------------------------
\institute{%
Dipartimento di Fisica, Universit\`a di Torino and INFN Sezione di Torino
\and
Departamento de F\`isica Te\`orica, Universitat de Val\`encia and IFIC, Universitat de Val\`encia-CSIC
\and
Istituto Nazionale di Fisica Nucleare, Sezione di Roma Tre
}
%----------------------------------------------------------------------------
\abstract{%
We present a new lattice determination of some of the parameters appearing both in the Operator Product Expansion (OPE) analysis of the inclusive semileptonic $B$-meson decays and in the Heavy Quark Expansion (HQE) of the pseudoscalar (PS) and vector (V) heavy-light meson masses. We perform a lattice QCD (LQCD) computation of PS and V heavy-light meson masses for heavy-quark masses $m_h$ in the range from $m_c^{\rm phys}$ to $\simeq 4m_b^{\rm phys}$. We employed the $N_f = 2+1+1$ gauge configurations of the European Twisted Mass Collaboration (ETMC) at three values of the lattice spacing $a \simeq (0.062, 0.082, 0.089)$ fm with pion masses in the range $M_\pi \simeq (210 - 450)$ MeV. The heavy-quark mass is simulated directly on the lattice up to $\simeq 3m_c^{\rm phys}$. The interpolation to the physical $m_b^{\rm phys}$ is performed using the ETMC ratio method and adopting the kinetic mass scheme. We obtain $m_b^{\rm kin}(1~\mbox{GeV}) = 4.61 (20)$ GeV ($\overline{m}_b(\overline{m}_b) = 4.26 (18)$ GeV in the $\overline{\rm MS}$ scheme). The lattice data are analyzed in terms of the HQE and the matrix elements of dimension-4 and dimension-5 operators are extracted with good precision, namely: $\overline{\Lambda} = 0.552 (26)$ GeV, $\mu_\pi^2 = 0.321 (32)$ GeV$^2$ and $\mu_G^2(m_b) = 0.253 (25)$ GeV$^2$. The data also allow for an estimate of the dimension-6 operator matrix elements.
}

%----------------------------------------------------------------------------
\maketitle
%----------------------------------------------------------------------------
\section{Introduction}
\label{intro}

The precise determination of the CKM element $V_{cb}$ is crucial in the search of new physics effects in rare decays (FCNC decays like $B_s \rightarrow\mu^+\mu^{-} \,,\,K^+ \rightarrow\pi^{+}\nu\overline{\nu}\,,\,K_L\rightarrow\pi_0\nu\overline{\nu}$ as well as $\epsilon_{K}$) and particularly interesting nowadays in view of the anomalies in $B \rightarrow D^{(*)}\tau\nu$. As is well known, there is a long-standing tension of about 3 standard deviations between the values of $V_{cb}$ obtained from inclusive and exclusive semileptonic B-meson decays. In the first case the OPE is usually adopted to describe the non-perturbative hadronic physics in terms of few parameters $\hat{\mu}_{i}^{2}$ , $\hat{\rho}_{i}^{3}$ extracted from experimental data on inclusive $B \to X_c \ell \nu_\ell$ decays together with $V_{cb}$ (see, e.g., Ref.~\cite{Gambino:2016jkc,Alberti:2014yda} and therein) :
\bea
\label{eq:Semilep}
    \Gamma\,[B\rightarrow  X_c \ell \nu_{\ell}] & = & \frac{G_F m_b^5}{192\pi^3}\,|V_{cb}|^2 \,g(r)A_{ew}\left[ 1 - \frac{\hat{\mu}_\pi^2-\hat{\mu}_G^2}{2m_b^2} - \frac{\hat{\rho}_D^3+\hat{\rho}_{LS}^3}{2m_b^3} -\frac{\hat{\mu}_G^2-\frac{\hat{\rho}_D^3+\hat{\rho}_{LS}^3}{m_b} }{m_b^2}\frac{2(1-r)^4}{g(r)}\right.  \nonumber  \\ 
    &+&\frac{d(r)}{g(r)}\frac{\hat{\rho}_D^3}{m_b^3}\left. + \sum_n \left( \frac{\alpha_s(m_b)}{\pi}\right)^n p_c^{(n)}(r,\mu)+\mathcal{O}\left(\frac{1}{m_b^4}\right)\right]\,,\hspace{1cm} r =\frac{m_c^2}{m_b^2} ~ .
\eea
In the second case the relevant hadronic inputs are the semileptonic form factors describing the $B \to D^* (D) \ell \nu_\ell$ decays, computed using LQCD simulations. Our aim is to present the lattice determination of some of the parameters appearing in the OPE analysis of the inclusive $B$-meson decays, obtained in Ref.~\cite{Gambino:2017vkx}.
Since the same parameters appear as coefficients of the HQE for the PS and V heavy-light meson masses, we study the heavy-quark mass dependence of two meson mass combinations, the spin averaged and the hyperfine splitting :
\bea
	\label{eq:Mass_Comb}
     	M_{av}(\widetilde{m}_h)  \equiv & (M_{PS}(\widetilde{m}_h) + 3 M_V(\widetilde{m}_h)) / 4  ~ ,
	    \qquad \Delta M(\widetilde{m}_h)  \equiv & M_V(\widetilde{m}_h) - M_{PS}(\widetilde{m}_h) ~ ,
\eea
where $\widetilde{m}_h = m_h^{kin}(\mu_{soft})$ is the renormalized heavy-quark mass in the kinetic scheme \cite{Bigi:1996si} at a soft cutoff $\mu_{soft}= 1$ GeV. 
The HQE of these quantities reads as 
\bea
      \label{eq:M_HQET_quartic}
       \frac{M_{av}(\widetilde{m}_h)}{\widetilde{m}_h} & = & 1 + \frac{\overline{\Lambda}}{\widetilde{m}_h} + 
            \frac{\mu_\pi^2}{2 \widetilde{m}_h^2} + \frac{\rho_D^3 - \rho_{\pi\pi}^3 - \rho_S^3}{4 \widetilde{m}_h^3} + 
            \frac{\sigma^4}{\widetilde{m}_h^4} ~ , \\[2mm]
     \label{eq:DM_HQET_quartic}
     \widetilde{m}_h \Delta M(\widetilde{m}_h) & = & \frac{2}{3} c_G(\widetilde{m}_h, \widetilde{m}_b) \mu_G^2(\widetilde{m}_b) + 
         \frac{\rho_{\pi G}^3 + \rho_A^3 - \rho_{LS}^3}{3 \widetilde{m}_h} + \frac{\Delta \sigma^4}{\widetilde{m}_h^2}~ ,
 \eea
where $\mu_i^2$, $\rho_i^2$ refer to matrix elements of asymptotically heavy mesons, directly related to $\hat{\mu}_i^2$, $\hat{\rho}_i^2$ in Eq.~(\ref{eq:Semilep}) that concerns matrix elements at the physical B-meson.
Generally only the charmed and beauty meson masses, $M_{D^{(*)}}$ and $M_{B^{(*)}}$, are used to constrain the HQE parameters (see Ref.~\cite{Gambino:2012rd}).
They could be constrained more effectively having meson masses with the heavy-quark mass between the physical charm and $b$-quark masses, $m_c^{\rm phys}$  and $m_b^{\rm phys}$, or even above $m_b^{\rm phys}$ as shown in Ref.~\cite{Kronfeld:2000gk}.  The main idea of this contribution is to exploit the ETMC ratio method~\cite{Blossier:2009hg} to employ LQCD as a virtual laboratory and compute these fictitious meson masses with good accuracy.

In Ref.~\cite{Gambino:2017vkx} the heavy-quark mass $\widetilde{m}_h$ defined in the kinetic scheme instead of the pole one, $m_h^{pole}$, was adopted. The main reason is that the relation between the pole mass and the bare lattice masses $\mu_h$, and therefore also the HQEs, suffers in perturbation theory from infrared renormalon ambiguities of order $O(\Lambda_{QCD})$ \cite{Bigi:1994em,Bigi:1996si,Beneke:1994sw,Luke:1994xd,Martinelli:1995vj}. The kinetic mass $\widetilde{m}_h$ is a short-distance mass that solves the problem by subtracting from the pole mass its infrared sensitive part \cite{Bigi:1996si,Czarnecki:1997sz}. The same scheme is often used in the analysis of the inclusive semileptonic $B$-meson decays relevant for the determination $V_{cb}$ \cite{Gambino:2016jkc,Alberti:2014yda}.

The relation between the simulated bare heavy-quark mass $a \mu_h$ and the kinetic mass $\widetilde{m}_h$ can be obtained in few steps.
Using the values of the lattice spacing and of the renormalization constant (RC) $Z_P$ determined in Ref.~\cite{Carrasco:2014cwa}, we first calculate $ \overline{m}_h(\mbox{2 GeV}) = (a \mu_h)/(Z_P ~ a)$ in the $\overline{\rm MS}$ scheme. Then we evolve the mass from $\mu = 2$ GeV to $\mu = \overline{m}_h$ using $\rm N^3LO$ perturbation theory \cite{Chetyrkin:1999pq} with four quark flavors ($n_\ell = 4$) and $\Lambda_{QCD}^{Nf = 4} = 297 (8)$ MeV \cite{PDG}. Finally, we make use of the relation between the kinetic mass $\widetilde{m}_h$ and the $\overline{\rm MS}$ mass $\overline{m}_h(\overline{m}_h)$, which is known up to two loops \cite{Gambino:2011cq} 
 \bea
       \widetilde{m}_h & \equiv & m_h^{pole} - \delta m_{h({\rm IR}) }^{pole}= \overline{m}_h(\overline{m}_h) \left\{ 1 + \frac{4}{3} \frac{\alpha_s(\overline{m}_h)}{\pi}  
                                           \left[1 - \frac{4}{3} x - \frac{1}{2} x^2 \right] + \left( \frac{\alpha_s(\overline{m}_h)}{\pi} \right)^2 
                                           \right. \nonumber \\
                                 & \cdot & \left. \left[ \frac{\beta_0}{24} (8 \pi^2 + 71) + \frac{35}{24} + \frac{\pi^2}{9} \mbox{ln}(2) - 
                                                 \frac{7 \pi^2}{12} -\frac{\zeta_3}{6} + \frac{4}{27} x \left( 24 \beta_0 \mbox{ln}(2x) - 64 \beta_0 + 6 \pi^2 - 39 \right) \right. \right.
                                           \nonumber \\
                                 & + & \left. \left. \frac{1}{18} x^2 \left( 24 \beta_0 \mbox{ln}(2x) - 52 \beta_0 + 6 \pi^2 - 23 \right)+ \frac{32}{27} x^3 -\frac{4}{9} x^4 \right] + {\cal{O}}(\alpha_s^3) \right\} ~ ,
       \label{eq:mkin}
 \eea
where $x \equiv \mu_{soft} / \overline{m}_h(\overline{m}_h)$, $\beta_0 = (33 - 2 n_\ell) / 12$ and $\zeta_3 \simeq 1.20206$. The conversion from the $\overline{\rm MS}$ to the kinetic scheme at the charm mass and specifically  the inclusion or not of the last two terms in Eq.~(\ref{eq:mkin}) give rise to what we will call the $()_{\rm conv}$ uncertainty in the error budget.

%----------------------------------------------------------------------------
\section{Extraction of the ground state meson masses and simulation details}
\label{sec-1}

\begin{figure}[htb!]
\centering
\includegraphics[width=6.5cm]{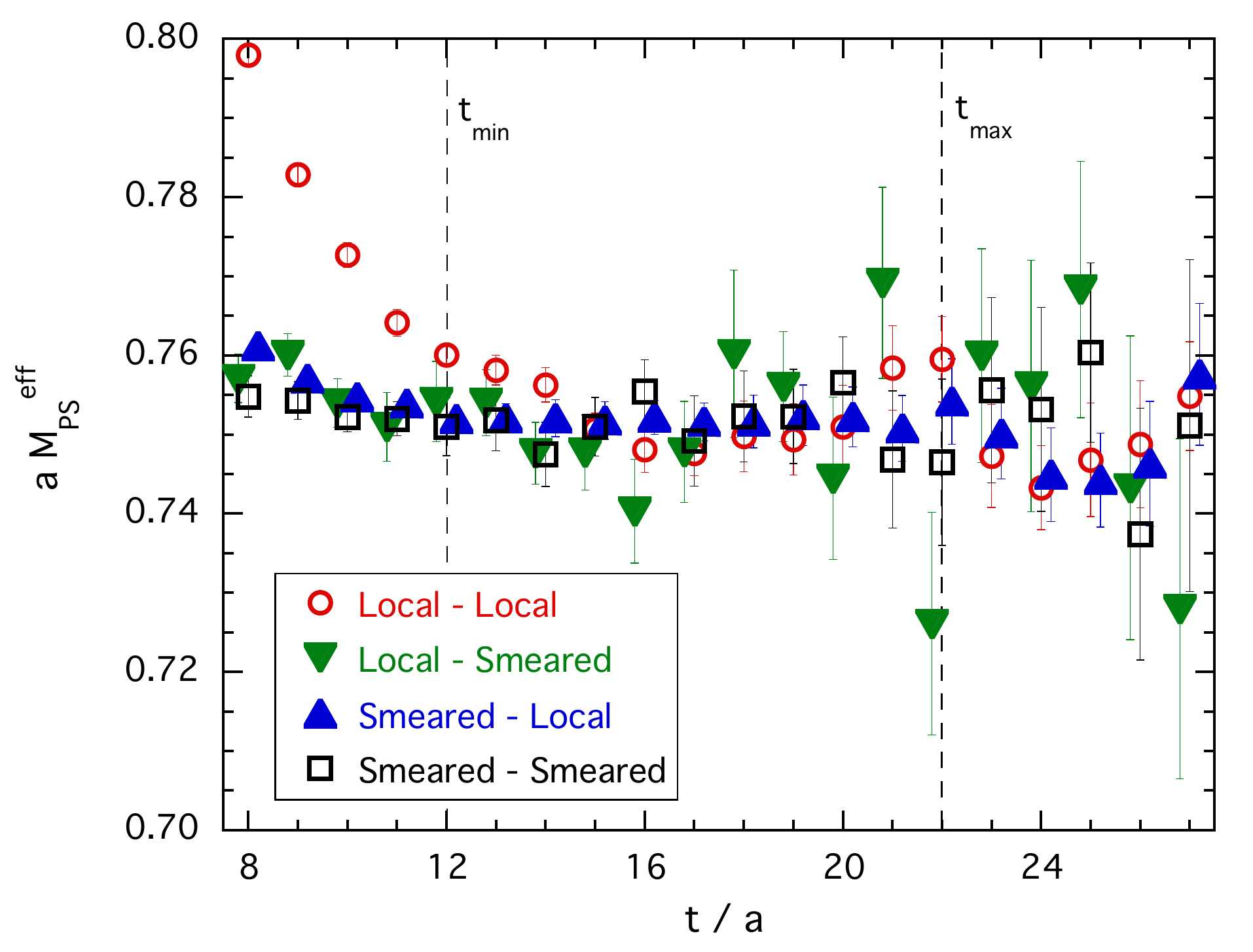}
\includegraphics[width=6.5cm]{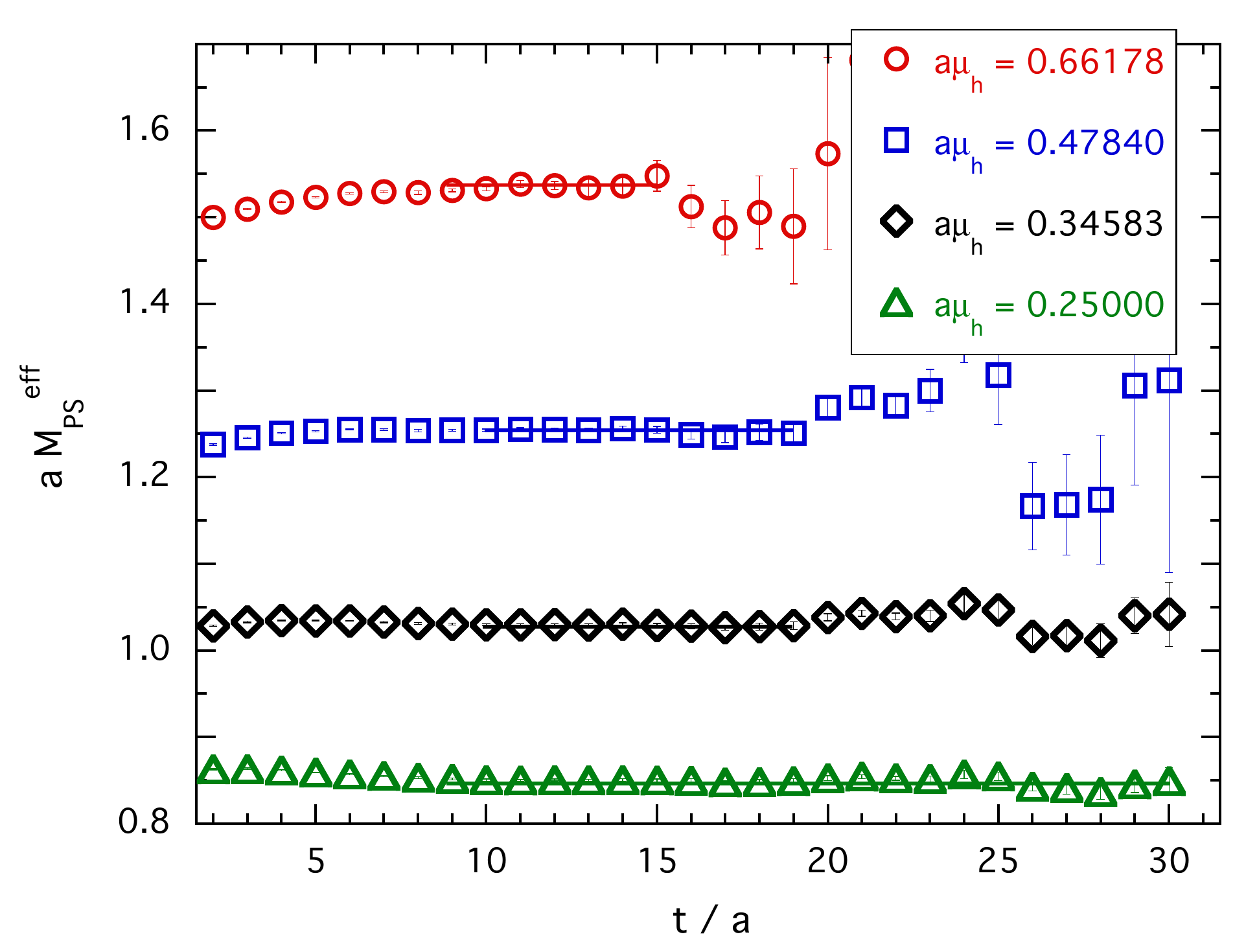}
\vspace{-0.2cm} 
\caption{\it \small Left panel: effective masses of the four correlators $LL, LS, SL, SS$, calculated for a ($c \ell$) meson using Eq.~(\ref{eq:Meff}) and corresponding to a pion mass equal to $\simeq 380$ MeV. Right panel: effective masses of the correlation $C_{PS(V)}^{SL}(t)$ calculated at various values of the bare heavy-quark masses $a\mu_{h}$ in lattice units.}
\label{fig:correlators}
\end{figure}

The PS(V) ground-state mass, $M_{PS(V)}$, can be extracted from the plateaux of the effective mass $M_{PS(V)}^{eff}(t)$, built out of a combination of the 2-point correlation function at various Euclidean times. In fact, at enough large values of $t \geq t_{\mathrm{min}}^{PS(V)}$ one has
 \be
     M_{PS(V)}^{eff}(t) \equiv \mbox{arcosh}\left[ \frac{C_{PS(V)}(t - 1) + C_{PS(V)}(t + 1)}{2 C_{PS(V)}(t)} \right] 
     ~ _{\overrightarrow{t \geq t_{\mathrm{min}}^{PS(V)}}} ~ M_{PS(V)} ~ .
     \label{eq:Meff}
 \ee
In order to obtain a faster suppression of the exited states, it is a common procedure to adopt Gaussian-smeared interpolating quark fields~\cite{Gusken:1989qx} both in the source and/or in the sink, namely $C_{PS(V)}^{LL}(t)$, $C_{PS(V)}^{LS}(t)$, $C_{PS(V)}^{SL}(t)$ and $C_{PS(V)}^{SS}(t)$, where $L$ and $S$ denote local and smeared operators, respectively. As illustrated in Fig.~\ref{fig:correlators}, the SL correlations allow for an earlier plateau together with an improved signal to noise ratio, and therefore they are used in our analysis to compute the ground-state masses.

The correlation functions used in this work refer to the gauge ensembles generated by ETMC with $N_f = 2 + 1 + 1$ dynamical quarks, which include in the sea, besides two light mass-degenerate quarks, also the strange and the charm quarks. In the ETMC set-up the gluon interactions are described by the Iwasaki action, while the fermions are regularized in the maximally twisted-mass Wilson formulation. The same ensembles are adopted in Refs.~\cite{Carrasco:2014cwa,Bussone:2016iua} to determine the up, down, strange, charm and bottom quark masses.  We have simulated three values of the valence charm quark mass, needed to interpolate smoothly in the physical charm region. The valence quark masses are in the ranges: $3 m_{ud}^{\rm phys} \lesssim m_\ell \lesssim 12 m_{ud}^{\rm phys}$ and $0.7 m_c^{\rm phys} \lesssim m_c \lesssim 1.1 m_c^{\rm phys}$. In order to extrapolate up to the $b$-quark sector we have also considered seven values of the valence heavy-quark mass, $m_h$, in the range $1.1 m_c^{\rm phys} \lesssim m_h \lesssim 3.3 m_c^{\rm phys} \approx 0.8 m_b^{\rm phys}$. The quality of the effective mass curves with the increase of the heavy quark mass is displayed in Fig.~\ref{fig:correlators}.
%=========
%=========
\vspace{-0.8cm} 
%----------------------------------------------------------------------------
\section{Analysis of the spin-averaged meson masses and determination of $m_b^{\rm phys}$}
\label{sec-2}

\begin{figure}[htb!]
       \centering
	\begin{minipage}[l]{10cm}
		\subfigure[\it]{\includegraphics[width=7cm]{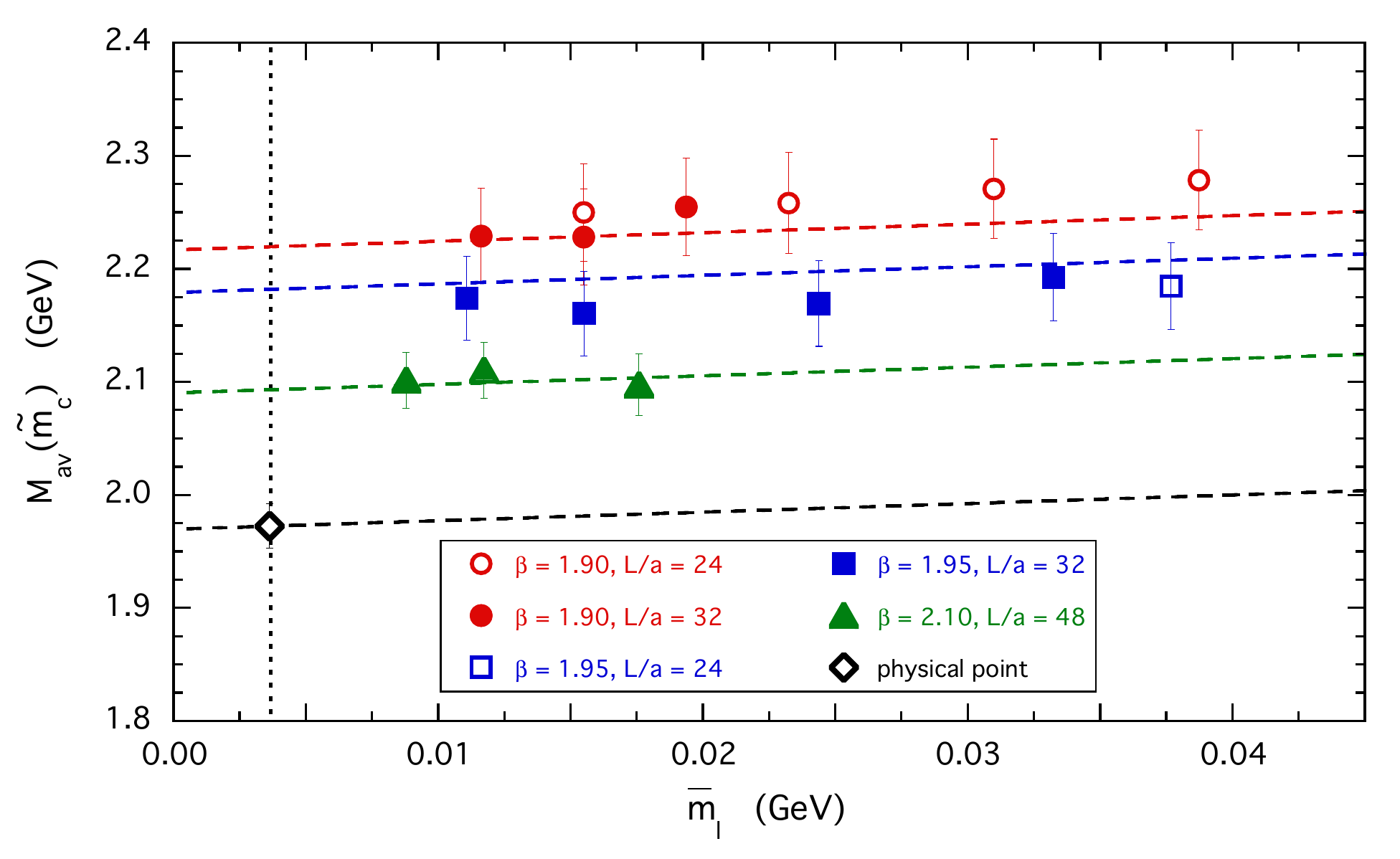}}\\[-3mm]
		\setcounter{subfigure}{2}
		\subfigure[\it]{\includegraphics[width=7cm]{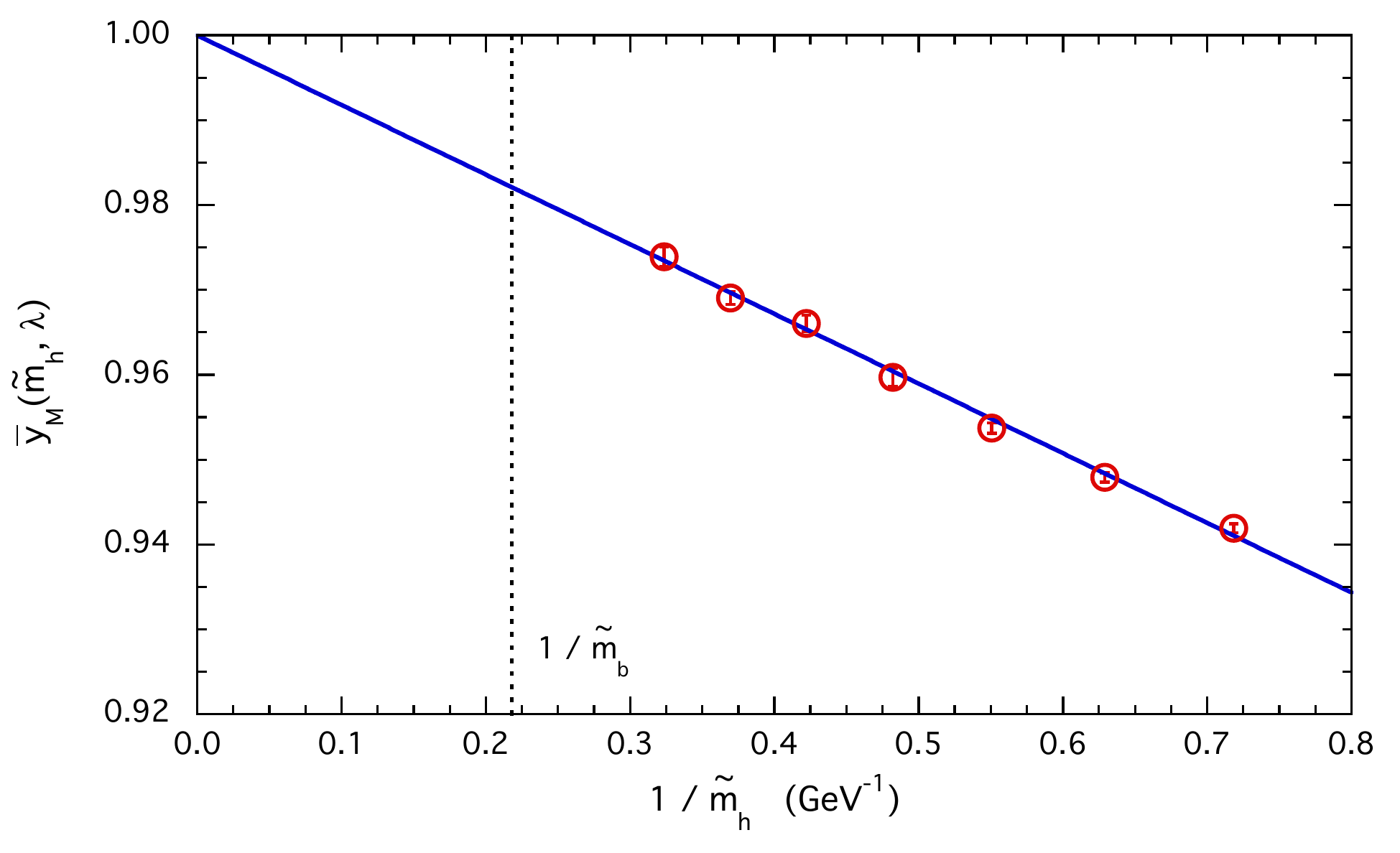}}
	\end{minipage}\hspace{-2.5cm}
	\begin{minipage}[r]{4cm}
		\includegraphics[width=4.4cm]{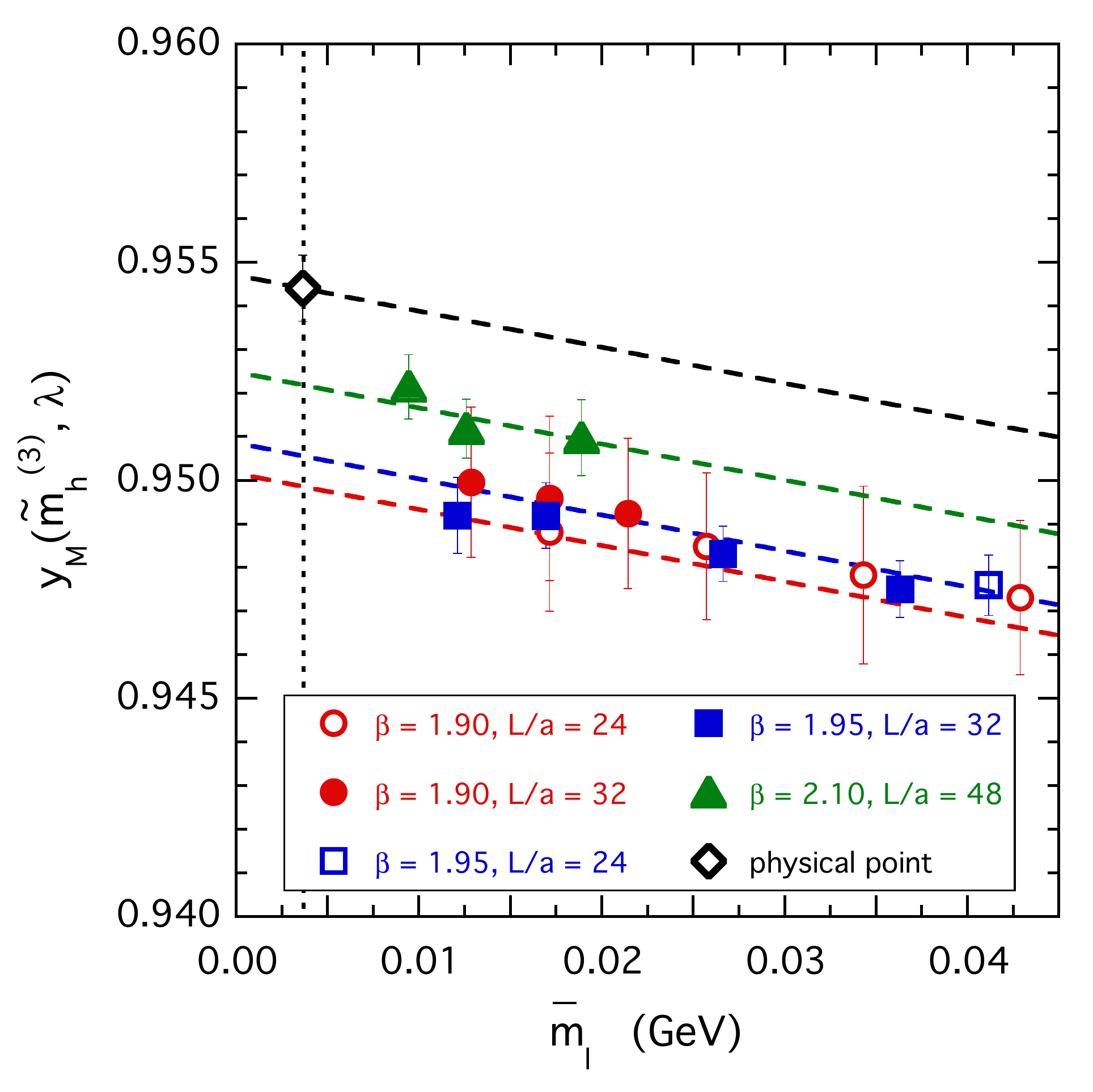}
		\setcounter{subfigure}{1}
		\subfigure[\it]{\includegraphics[width=4.4cm]{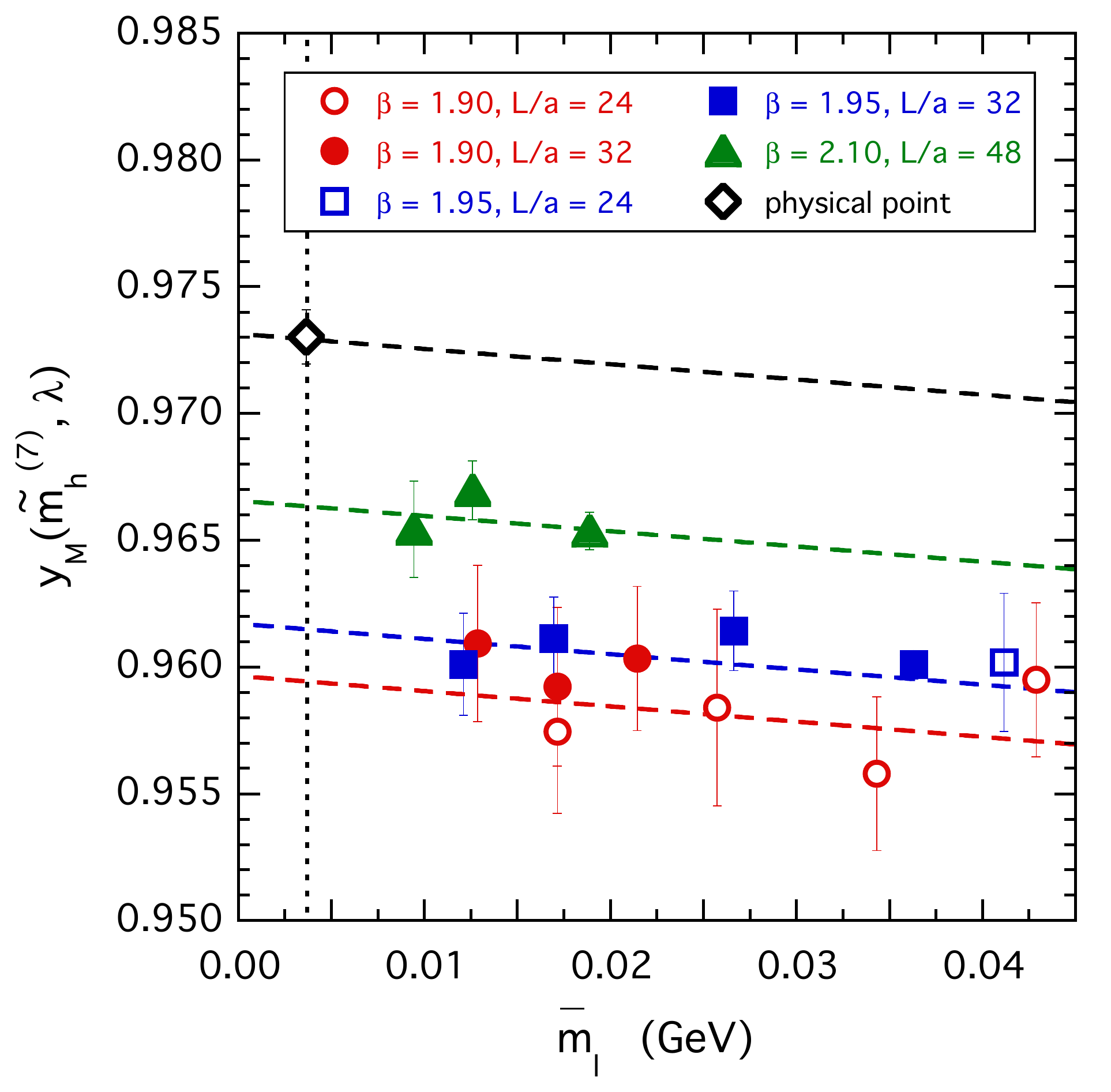}}
	\end{minipage}
	\vspace{-0.4cm}
	\caption{ (a) {\it \small Combined chiral and continuum limit of the quantity $M_{av}(\widetilde{m}_h^{(1)}) = M_{av}(\widetilde{m}_c)$ versus the (renormalized) light-quark mass $\overline{m}_\ell(\mbox{2 GeV})$.} (b) {\it \small Combined chiral and continuum limit of the ratios $y_{M}$ for $\widetilde{m}_h^{(4)}$ (upper panel) and $\widetilde{m}_h^{(8)}$ (lower panel) versus the (renormalized) light-quark mass $\overline{m}_\ell(\mbox{2 GeV})$.} (c) {\it \small Ratios $\overline{y}_{\Delta M}(\widetilde{m}_h, \lambda)$ versus the inverse heavy-quark mass $1 / \widetilde{m}_h$. The solid line is the result of the fit (\ref{eq:yM_fit}) with $\Delta \epsilon_2 = 0$, taking into account the correlations among the lattice points.}
	\label{fig:Mav}}
\end{figure}

We start by applying the ETMC ratio method~\cite{Blossier:2009hg} to the quantity $M_{av}(\widetilde{m}_h)$ in Eq.~(\ref{eq:Mass_Comb}).
To this end we construct a sequence of heavy-quark masses $\lbrace\widetilde{m}_h^{(n)}\rbrace$ with a common fixed ratio $\lambda$: $\widetilde{m}_h^{(n)} = \lambda \widetilde{m}_h^{(n - 1)} $. The series starts at the physical charm quark mass $\widetilde{m}_h^{(1)} = \widetilde{m}_c = 1.219\, (41) ~ (40)_{\rm conv}$ GeV corresponding to  $\overline{m}_c(\mbox{2 GeV}) = 1.176 (36)$ GeV, obtained in Ref.~\cite{Carrasco:2014cwa}. For each gauge ensemble the quantity $M_{av}(\widetilde{m}_c)$ can be computed by a smooth interpolation of the results in the charm region and a subsequent extrapolation to the physical pion mass and to the continuum limit using a simple, combined linear fit in both $\overline{m}_\ell$ and $a^2$ as shown in Fig.~\ref{fig:Mav}(a). We get $M_{av}^{\rm phys}(\widetilde{m}_c) = 1.967 (25)$ GeV, which agrees with the experimental value $(M_D + 3 M_{D^*}) / 4 = 1.973$ GeV from PDG \cite{PDG} as well as with the result $M_{av}^{\rm phys}(\widetilde{m}_c) = 1.975 (11)$ GeV based on the direct investigation of the V to PS meson mass ratios of Ref.~\cite{Lubicz:2017asp}.
%========
%========
Analogously, for each gauge ensemble the quantities $M_{av}(\widetilde{m}_h^{(n)})$ with $n = 2, 3, ...$ can be evaluated by interpolating the results corresponding to the subset of the bare heavy-quark masses.
Then, we can construct the following ratios
 \be
      y_M(\widetilde{m}_h^{(n)}, \lambda) = \frac{M_{av}(\widetilde{m}_h^{(n)})}{M_{av}(\widetilde{m}_h^{(n - 1)})} 
                                                                   \frac{\widetilde{m}_h^{(n - 1)}}{\widetilde{m}_h^{(n)}}
                                                                = \lambda^{-1} \frac{M_{av}(\widetilde{m}_h^{(n)})}{M_{av}(\widetilde{m}_h^{(n - 1)})} \qquad (n = 2, 3, ...) ~ ,
      \label{eq:yM}
 \ee
which have the static limit $\mbox{lim}_{\widetilde{m}_h \to \infty} ~ y_M(\widetilde{m}_h, \lambda) = 1$ (see Eq.~(\ref{eq:M_HQET_quartic})).
Each ratio is extrapolated to the physical pion mass and to the continuum limit using again a combined linear fit in both $\overline{m}_\ell$ and $a^2$, obtaining values denoted by $\overline{y}_M(\widetilde{m}_h^{(n)}, \lambda)$. Considering the ratios (\ref{eq:yM}) has the advantage that discretization effects are suppressed, as illustrated in Fig.~\ref{fig:Mav}(b). Thus the $\widetilde{m}_h$-dependence of $\overline{y}_M$ can be described as a series expansion in terms of $1/\widetilde{m}_h$, namely
 \be
     \overline{y}_M(\widetilde{m}_h, \lambda) = 1 +\epsilon_1 / \widetilde{m}_h +  \epsilon_2 / \widetilde{m}_h^2 + 
                                                                         {\cal{O}}\left( 1/ \widetilde{m}_h^3 \right) ~ .
     \label{eq:yM_fit}
 \ee
In Fig.~\ref{fig:Mav}(c) it can be seen that a linear fit with $\epsilon_2 = 0$ is sufficient to fit the data, adopting a correlated $\chi^2$-minimization procedure. 
Finally, the chain equation
 \be
     M_{av}(\widetilde{m}_b) \equiv M_{av}(\widetilde{m}_h^{(K+1)}) = \lambda^{-K} M_{av}(\widetilde{m}_c) \,\,\overline{y}_M(\widetilde{m}_h^{(2)}, \lambda) \,\,~ \overline{y}_M(\widetilde{m}_h^{(3)}, \lambda) ~ ... ~ \overline{y}_M(\widetilde{m}_h^{(K+1)}, \lambda) ~ ,
     \label{eq:chain_M}
 \ee
allows to determine the $b$-quark mass $\widetilde{m}_b$ in an iterative way, requiring that, tuning the parameter $\lambda$, after $K$ steps the quantity $M_{av}(\widetilde{m}_b)$ matches the experimental value $(M_B + 3 M_{B^*}) / 4 = 5.314$ GeV \cite{PDG}. Then the $b$-quark mass $\widetilde{m}_b$ is directly given by $\widetilde{m}_b = \widetilde{m}_h^{(K+1)}=\lambda^K ~ \widetilde{m}_c$.
Adopting $K = 10$ we find $\lambda = 1.1422 (10)$, which yields 
 \be
     \widetilde{m}_b = 4.605 ~ (120)_{\rm stat} ~ (57)_{\rm syst} ~  (150)_{\rm conv}~ \mbox{GeV}  ~ .
     \label{eq:mb_kin}
\ee
In the $\overline{\rm MS}$ scheme the result (\ref{eq:mb_kin}) corresponds to $\overline{m}_b(\overline{m}_b) = 4.257$ $(120)$ GeV, which is well compatible with the ETMC determination $\overline{m}_b(\overline{m}_b) = 4.26$ $(10)$ GeV given in Ref.~\cite{Bussone:2016iua} and consistent with other lattice determinations within one standard deviation (see, e.g., the FLAG review~\cite{FLAG}).

\begin{figure}[htb!]
       \centering
	\includegraphics[width=9cm]{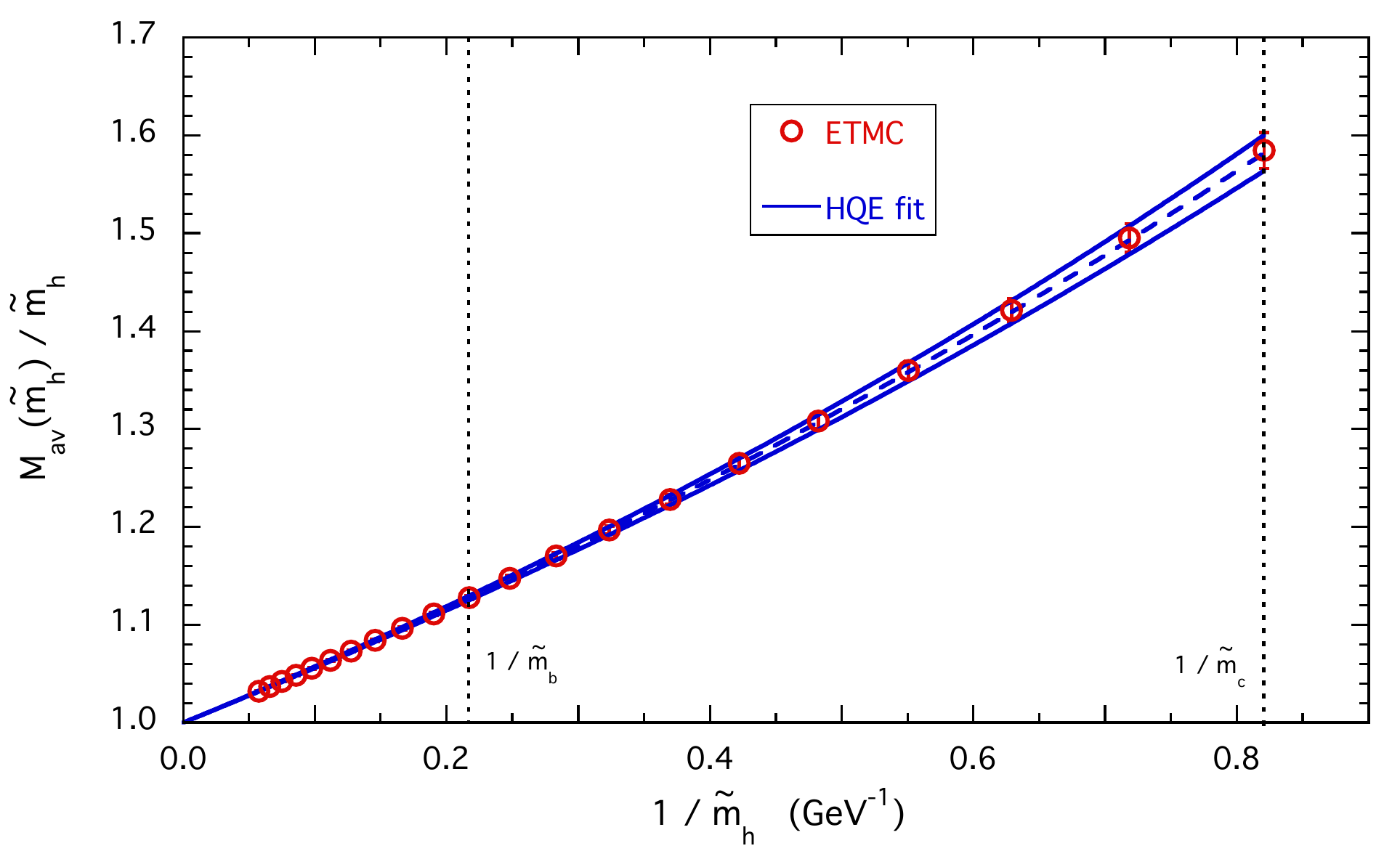}
	\vspace{-0.4cm}
	\caption{\it \small Lattice data for the quantity $M_{av}(\widetilde{m}_h) / \widetilde{m}_h$ versus the inverse heavy-quark mass $\widetilde{m}_h$. The dashed and solid lines are the results of the HQE fit (\ref{eq:M_HQET_quartic}) taking into account the correlations among the lattice points.
	\label{fig:Mav_HQET}}
\end{figure}

The chain equation (\ref{eq:chain_M}) can be easily extended beyond the physical $b$-quark point, $n > K+1$, using the fitting function (\ref{eq:yM_fit}) with $\epsilon_2 = 0$.
In the case of the spin-averaged meson mass one obtains
 \be
      \frac{M_{av}(\widetilde{m}_h^{(n)})}{\widetilde{m}_h^{(n)}} =\frac{M_{av}(\widetilde{m}_c)}{\widetilde{m}_c} ~  \prod_{i = 2}^{n}
                    \overline{y}_M(\widetilde{m}_h^{(i)}, \lambda) ~ = \frac{M_{av}(\widetilde{m}_c)}{\widetilde{m}_c} ~  \prod_{i = 2}^{n}
                    \left[ 1 + \frac{\epsilon_1}{\lambda^{i-1} \widetilde{m}_c} \right] ~ ,
     \label{eq:Mavn}
 \ee
We have evaluated Eqs.~(\ref{eq:Mavn}) for $n \lesssim 20$, i.e.~for heavy-quark masses up to $\widetilde{m}_h \simeq 4 \widetilde{m}_b$. The results, shown in Fig.~\ref{fig:Mav_HQET}, exhibit uncertainties at the level of $\simeq (1-2) \%$, vanishing in the static limit. 
%========
%========
Taking into account the correlations between the lattice data, we have performed the HQE fit (\ref{eq:M_HQET_quartic}), obtaining
\bea
    \label{eq:Lambda}
     \overline{\Lambda} & = & 0.552 ~ (13)_{\rm stat+syst} ~ (22)_{\rm conv} ~ \mbox{GeV} ~ , \\
    \label{eq:kinetic}
     \mu_\pi^2 & = & 0.321 ~ (17)_{\rm stat+syst} ~ (27)_{\rm conv} ~ \mbox{GeV}^2 , \\
    \label{eq:rho_M}
     \rho_D^3 - \rho_{\pi\pi}^3 - \rho_S^3 & = & 0.153 ~ (30)_{\rm stat+syst} ~ (17)_{\rm conv} ~ \mbox{GeV}^3 , \\
     \sigma^4 & = &  0.0071 ~ (55)_{\rm stat+syst}  ~ \mbox{GeV}^4 ~ ,  
 \eea
where the systematic uncertainty includes both the spread of the results of different fits ($\widetilde{m}_{h}\geq \widetilde{m}_{c}$ with $\sigma^4 = 0$, $\widetilde{m}_{h}\geq 2\widetilde{m}_{c}$ with $\sigma^4 = 0$, $\widetilde{m}_{h}\geq \widetilde{m}_{c}$ with $\sigma^4 \neq 0$) and the bootstrap samplings of the inputs parameters (lattice spacing, RC $Z_P$ and physical quark masses) corresponding to the eight branches of the analysis of Ref.~\cite{Carrasco:2014cwa}. 

%----------------------------------------------------------------------------
\section{Analysis of the hyperfine meson mass splittings}
\label{sec-3}

\begin{figure}[htb!]
	\centering
	\begin{minipage}[l]{10cm}
		\subfigure[\it]{\includegraphics[width=7cm]{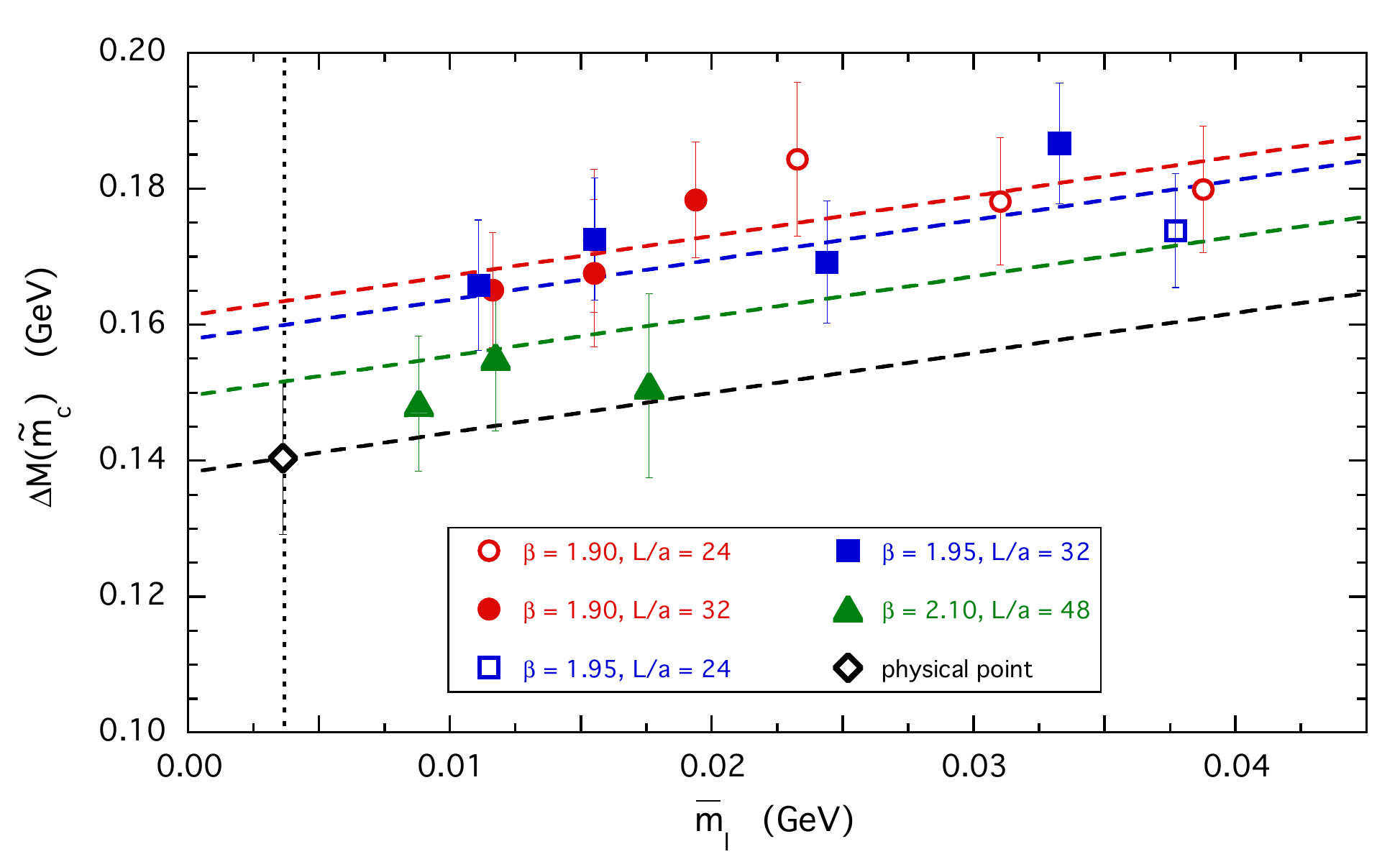}}\\[-4mm]
		\setcounter{subfigure}{2}
		\subfigure[\it]{\includegraphics[width=7cm]{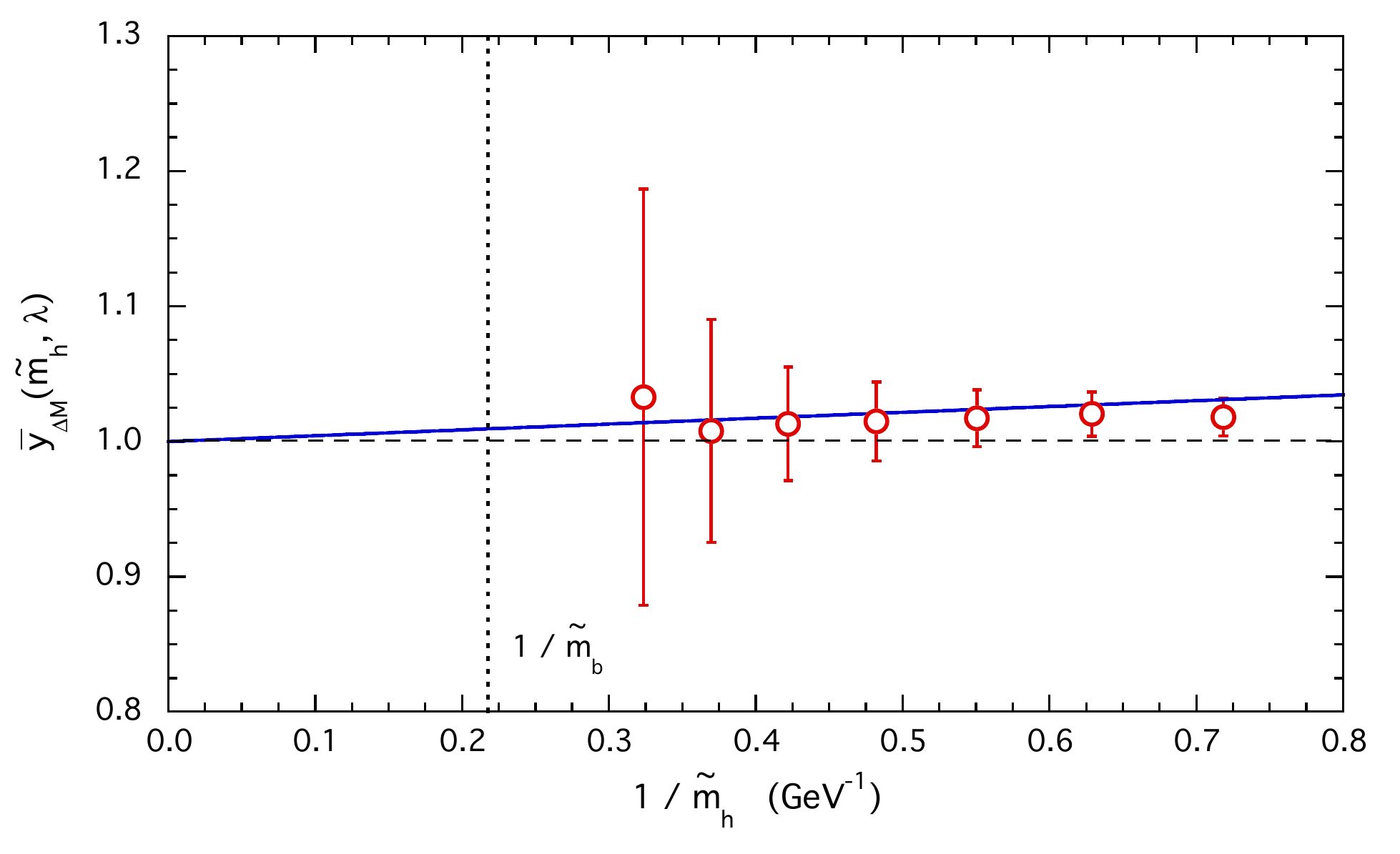}}
	\end{minipage}\hspace{-2.5cm}
	\begin{minipage}[r]{4cm}
		\includegraphics[width=4.4cm]{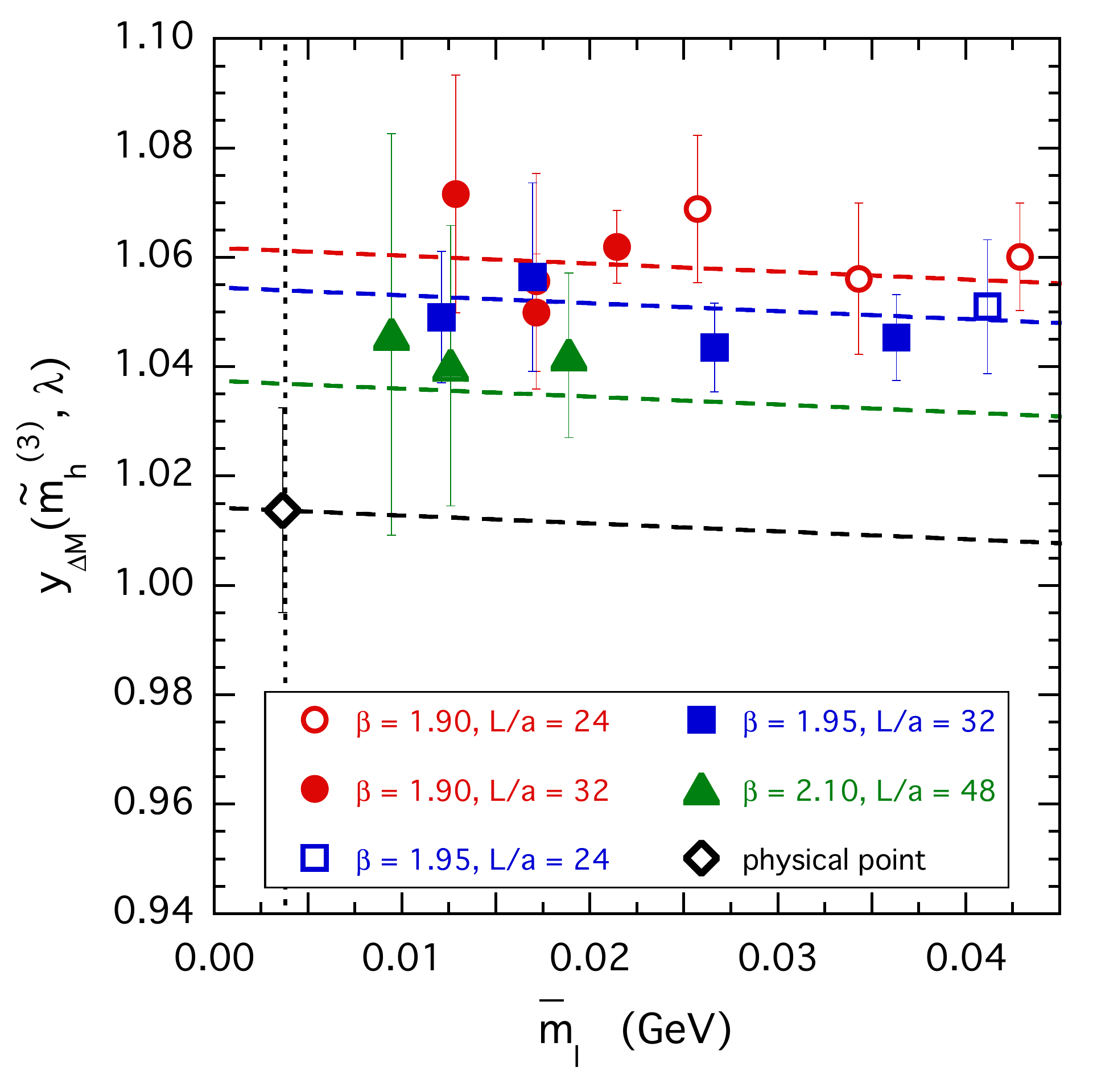}
		\setcounter{subfigure}{1}
		\subfigure[\it]{\includegraphics[width=4.4cm]{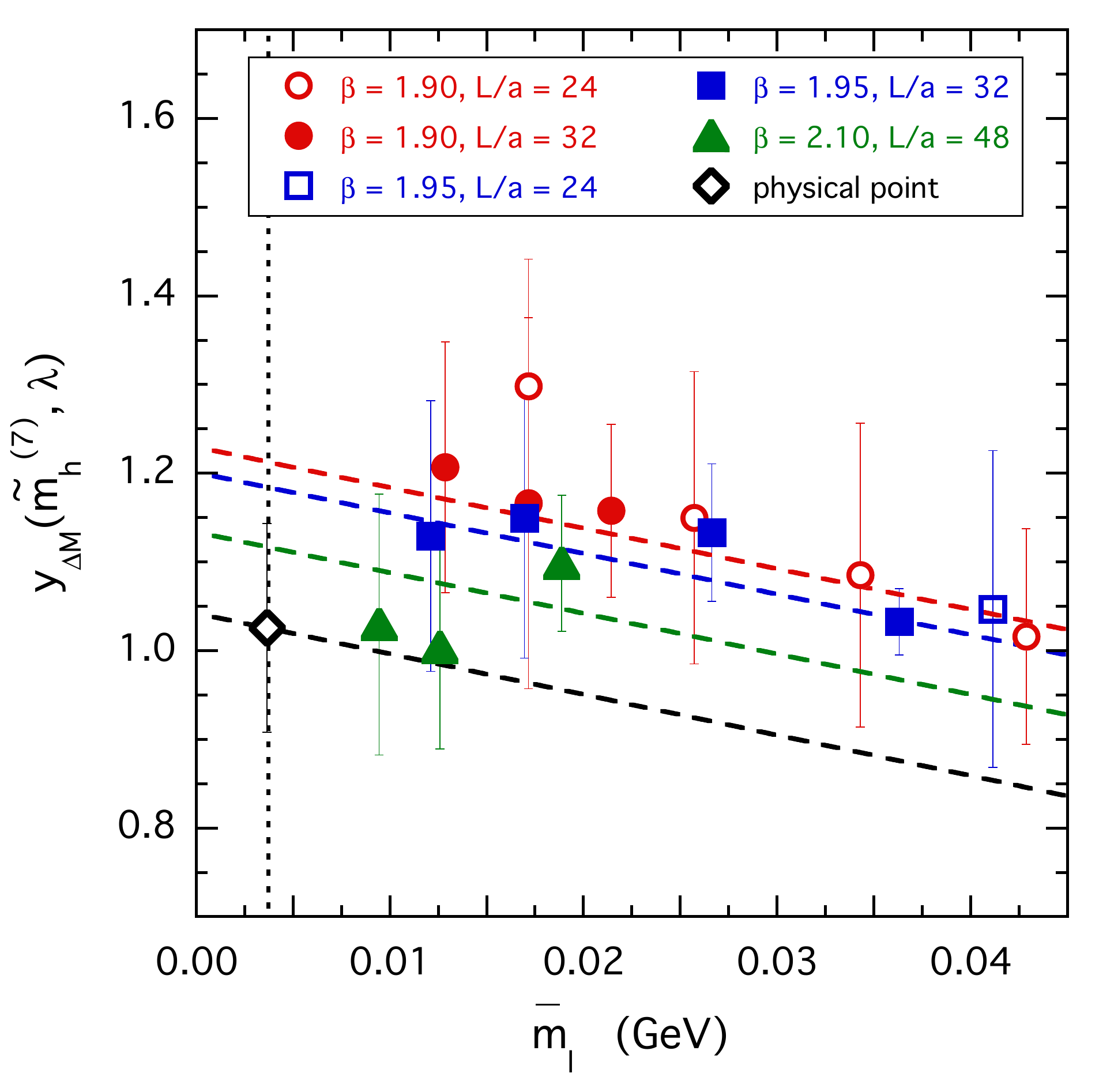}}
	\end{minipage}
	\vspace{-0.25cm}
	\caption{ (a) {\it \small Combined chiral and continuum limit of the quantity $\Delta M(\widetilde{m}_h^{(1)}) = \Delta M(\widetilde{m}_c)$ versus the (renormalized) light-quark mass $\overline{m}_\ell(\mbox{2 GeV})$.} (b) {\it \small Combined chiral and continuum limit of the ratios $y_{\Delta M}$ for $\widetilde{m}_h^{(4)}$ (upper panel) and $\widetilde{m}_h^{(8)}$ (lower panel) versus the (renormalized) light-quark mass $\overline{m}_\ell(\mbox{2 GeV})$.} (c) {\it \small Ratios $\overline{y}_{\Delta M}(\widetilde{m}_h, \lambda)$ versus the inverse heavy-quark mass $1 / \widetilde{m}_h$. The solid line is the result of the linear fit (\ref{eq:yM_fit}) ($\Delta \epsilon_2 = 0$), taking into account the correlations among the lattice points.}
	\label{fig:DM}}
\end{figure}

In this Section we apply the ratio method to the hyperfine meson mass splitting $\Delta M(\widetilde{m}_h)$ in Eq.~(\ref{eq:Mass_Comb}). As in the case of the spin-averaged case, for each gauge ensemble the quantity $\Delta M(\widetilde{m}_c)$ is computed by interpolating first to the charm region and then extrapolating to the physical pion mass and to the continuum limit using a combined linear fit in both $\overline{m}_\ell$ and $a^2$, as illustrated in Fig.~\ref{fig:DM}(a). We get $\Delta M^{\rm phys}(\widetilde{m}_c) = 140 (11)$ MeV, which nicely agrees with the experimental value $M_{D^*} - M_D = 141.4$ MeV from PDG \cite{PDG} as well as with the result $M_{D^*} - M_D = 144 (15)$ MeV obtained in Ref.~\cite{Lubicz:2017asp} from a direct investigation of the $D^*$- to $D$-meson mass ratio. Analogously, for each gauge ensemble the quantities $\Delta M(\widetilde{m}_h^{(n)})$ with $n = 2, 3, ...$ are evaluated by interpolating the results corresponding to the bare heavy-quark masses.
We now consider the following ratios
 \be
     y_{\Delta M}(\widetilde{m}_h^{(n)}, \lambda) =  \frac{\widetilde{m}_h^{(n )}}{\widetilde{m}_h^{(n - 1)}}
         \frac{\Delta M(\widetilde{m}_h^{(n)})}{\Delta M(\widetilde{m}_h^{(n - 1)})} 
         \frac{c_G(\widetilde{m}_h^{(n - 1)}, \widetilde{m}_b)}{c_G(\widetilde{m}_h^{(n)}, \widetilde{m}_b)} =
         \lambda \frac{\Delta M(\widetilde{m}_h^{(n)})}{\Delta M(\widetilde{m}_h^{(n - 1)})} 
         \frac{c_G(\widetilde{m}_h^{(n - 1)}, \widetilde{m}_b)}{c_G(\widetilde{m}_h^{(n)}, \widetilde{m}_b)} ~ ,
     \label{eq:yDM}
 \ee
which have the static limit $\mbox{lim}_{\widetilde{m}_h \to \infty} ~ y_{\Delta M}(\widetilde{m}_h, \lambda) = 1$ (see Eq.~(\ref{eq:DM_HQET_quartic})).
In Eq.~(\ref{eq:yDM}) $c_G(\widetilde{m}_h, \widetilde{m}_b)$ is the short-distance Wilson coefficient that multiplies the matrix element of the HQET chromomagnetic operator. It can be factorized in three parts: $c_G = \overline{c}_G \cdot {\mathcal{R}} \cdot (\widetilde{m}_h/m_h^{pole})$ where the conversion coefficient $\overline{c}_G$ is known up to three loops \cite{Grozin:2007fh} and ${\mathcal{R}}$ is the evolution factor (see Ref.~\cite{Gambino:2017vkx})
 \be
    c_G (\widetilde{m}_h, \widetilde{m}_b) = \left[ 1 + \frac{13}{6} ~ \frac{\alpha_s(\widetilde{m}_h)}{\pi} + 
        (11\beta_0 - 10 ) ~  \left( \frac{\alpha_s(\widetilde{m}_h)}{\pi} \right)^2 \right] 
        \frac{R(\widetilde{m}_h)}{R(\widetilde{m}_b)} \left( \frac{\alpha_s(\widetilde{m}_h)}
        {\alpha_s(\widetilde{m}_b)} \right)^{\frac{\gamma_0}{2 \beta_0}} 
        \frac{\widetilde{m}_h}{m_h^{pole}} ~ .
\ee
The ratios (\ref{eq:yDM}) are extrapolated to the physical pion mass and to the continuum limit using a combined linear fit in both $\overline{m}_\ell$ and $a^2$ (see Fig.~\ref{fig:DM}(b)), obtaining values denoted by $\overline{y}_{\Delta M}(\widetilde{m}_h^{(n)}, \lambda)$. The $\widetilde{m}_h$-dependence of $\overline{y}_{\Delta M}$ can be described as a series expansion in terms of $1/\widetilde{m}_h$ analogous to Eq.~(\ref{eq:yM_fit}) with $\epsilon \rightarrow \Delta\epsilon$. Again, a linear fit ($\Delta\epsilon_2 = 0$) is sufficient to reproduce the data, as shown in Fig.~\ref{fig:DM}(c).

\begin{figure}[htb!]
	\centering
	\includegraphics[width=9cm]{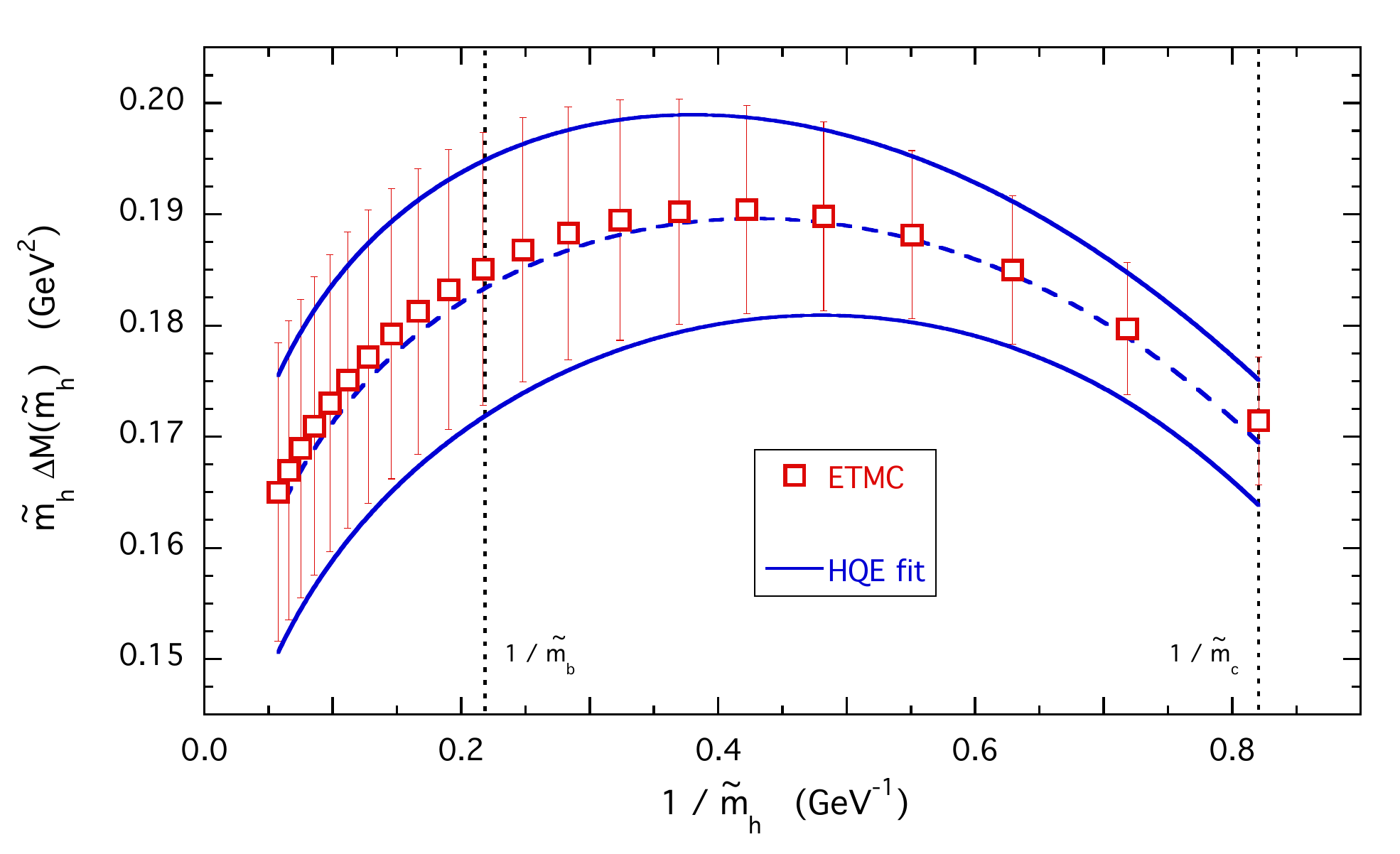}
	\vspace{-0.2cm}
	\caption{\it \small Lattice data for the quantity $\widetilde{m}_h^{(n)} \Delta M(\widetilde{m}_h^{(n)})$ versus the inverse heavy-quark mass $\widetilde{m}_h$. The dashed and solid lines are the results of the HQE fit (\ref{eq:DM_HQET_quartic}), taking into account the correlations among the lattice points.
	\label{fig:DM_HQET}}
\end{figure}

Using a chain equation analogous to Eq.~(\ref{eq:chain_M}) but expressed in terms of the ratios (\ref{eq:yDM}) and adopting the values of the parameters $\lambda$ and $K$ determined in the previous Section to reach the physical $b$-quark mass (\ref{eq:mb_kin}), we get for the hyperfine $B$-meson mass splitting the result $\Delta M(\widetilde{m}_b) = M_{B^*} - M_B = 40.2 (2.1)$ MeV, which is slightly below the experimental value $M_{B^*} - M_B = 45.42 (26)$ MeV \cite{PDG}, but improves the result $M_{B^*} - M_B = 41.2 (7.4)$ MeV of Ref.~\cite{Lubicz:2017asp}, based on the direct investigation of the V to PS meson mass ratios.
Going beyond the physical $b$-quark point, we obtain
\bea
      \widetilde{m}_h^{(n)} \frac{\Delta M(\widetilde{m}_h^{(n)})}{c_G(\widetilde{m}_h^{(n)}, \widetilde{m}_b)} = 
          \widetilde{m}_c \frac{\Delta M(\widetilde{m}_c)}{c_G(\widetilde{m}_c, \widetilde{m}_b)} ~  \prod_{i = 2}^{n}
          \overline{y}_{\Delta M}(\widetilde{m}_h^{(i)}, \lambda) = \widetilde{m}_c \frac{\Delta M(\widetilde{m}_c)}{c_G(\widetilde{m}_c, \widetilde{m}_b)} ~  \prod_{i = 2}^{n} 
          \left[ 1 + \frac{\Delta \epsilon_1}{\lambda^{i-1} \widetilde{m}_c} \right] 
     \label{eq:DMn}
 \eea
for $K+1 \leq n \lesssim 20$. Taking into account the correlations between lattice data, we have performed the HQE fit ansatze (\ref{eq:DM_HQET_quartic}), as shown in Fig.~\ref{fig:DM_HQET}. Our final results for the HQE parameters are:
\bea
    \label{eq:cromomagnetic}
     \mu_G^2(\widetilde{m}_{b}) & = & 0.253 ~ (21)_{\rm stat+syst} ~ (13)_{\rm conv} ~ \mbox{GeV}^2 , \\
    \label{eq:rho_DM}
     \rho_{\pi G}^3 + \rho_A^3 - \rho_{LS}^3 & = & -0.158 ~ (71)_{\rm stat+syst} ~ (45)_{\rm conv} ~ \mbox{GeV}^3,\\
     \Delta \sigma^4 & = & 0.0092 ~ (60)_{\rm stat+syst} ~ \mbox{GeV}^4 ~ . 
 \eea
%========
%========

%----------------------------------------------------------------------------
\section{Conclusions}

We have presented a precise lattice computation of the pseudoscalar and vector meson masses up to heavy-quark masses $m_h \simeq 4 m_b^{\rm phys}$. They allow for a new, precise unquenched lattice determination of the dimension-5 and dimension-6 HQE parameters relevant for the determination of the CKM entry $V_{cb}$ from the analysis of the inclusive semileptonic $B$-meson decays. We have adopted the ETMC gauge configurations with $N_f=2+1+1$ dynamical quarks at $a\simeq(0.06-0.09)$fm and $M_{\pi}\simeq(210-450)$MeV. The heavy-quark mass has been simulated directly on the lattice up to $\simeq 3$ times the physical charm mass. The interpolation to the physical $b$-quark mass has been performed using the ETMC ratio method adopting the kinetic mass scheme in order to work with a short-distance mass free from renormalon ambiguities. Our complete set of results (see Eqs.~(\ref{eq:M_HQET_quartic}-\ref{eq:DM_HQET_quartic})) are: $\widetilde{m}_c =  1.219 ~ (57) ~ \mbox{GeV}$, $\widetilde{m}_b  = 4.605 ~ (201) ~ \mbox{GeV}$, $\overline{\Lambda} = 0.552 ~ (26) ~\mbox{GeV}$, $\mu_\pi^2 = 0.321 ~ (32)~\mbox{GeV}^2$, $\mu_G^2(m_b) = 0.253 ~ (25)~\mbox{GeV}^2$, $\rho_D^3 - \rho_{\pi \pi}^3 - \rho_S^3 = 0.153 ~ (34) ~\mbox{GeV}^3$, $\rho_{\pi G}^3 + \rho_A^3 - \rho_{LS}^3 = -0.158 ~ (84) ~\mbox{GeV}^3$, $\sigma^4 =  0.0071 ~ (55)~ \mbox{GeV}^4$ and $\Delta \sigma^4 = 0.0092 ~ (60)~ \mbox{GeV}^4$ .

\bibliography{lattice2017}

\end{document}